\documentclass{sig-alternate}
\usepackage{amssymb}
\usepackage{amsmath}
\usepackage{color}
\usepackage[colorlinks=true,citecolor=blue,linkcolor=blue,urlcolor=blue]{hyperref} 
\usepackage{graphicx}
\usepackage{caption}
\DeclareCaptionType{copyrightbox}
\usepackage{subcaption}
\usepackage{wrapfig}

\begin{document}
\title{RAProp: Ranking Tweets by Exploiting the Tweet/User/Web Ecosystem and Inter-Tweet Agreement}

\author{Srijith Ravikumar$^\dag$,\ Kartik Talamadupula$^\dag$,\ Raju Balakrishnan$^\S$,\ Subbarao Kambhampati$^\dag$ \\  
  \and
  $^\dag$Dept. of Computer Science and Engg.\\
  Arizona State University\\
  Tempe AZ 85287\\
   \begin{normalsize}{\tt \{srijith,krt,rao\}} @ {\tt
       asu.edu}\end{normalsize} \and
   $^\S$Groupon, Inc.\\
   3101 Park Blvd\\
   Palo Alto CA 94306\\
   \begin{normalsize}{\tt raju} @ {\tt
       groupon.com}\end{normalsize}}

\maketitle
\begin{abstract}
The increasing popularity of Twitter renders improved trustworthiness
and relevance assessment of tweets much more important for
search. However, given the limitations on the size of tweets, it is
hard to extract measures for ranking from the tweets' content
alone. We present a novel ranking method, called \emph{RAProp}, which
 combines two orthogonal measures of relevance and
trustworthiness of a tweet. The first, called Feature
Score, measures the trustworthiness of the {\em source} of the tweet. This
is done by extracting features from a 3-layer twitter ecosystem,
consisting of users, tweets and the pages referred to in the
tweets. The second measure, called agreement analysis, estimates the
trustworthiness of the {\em content} of the tweet, by analyzing how
and whether the content is independently corroborated by other
tweets. We view the candidate result set of tweets as the vertices of
a graph, with the edges measuring the
estimated agreement between each pair of tweets. The feature score is
propagated over this agreement graph to compute the top-k tweets that
have both trustworthy sources and independent corroboration.
The evaluation of our method on 16~million tweets from the TREC 2011 Microblog Dataset shows that for top-$30$ precision we achieve $53\%$ higher than current best performing method on the Dataset and over $300\%$ over current Twitter Search. We also present a detailed internal empirical evaluation of {\em RAProp} in comparison
to several alternative approaches proposed by us.
\end{abstract}

\section{Introduction}
Twitter, the popular microblogging service, is increasingly being
looked upon as a source of the latest news and trends. The open nature
of the platform, as well as the lack of restrictions on who can post
information on it, leads to fast dissemination of all kinds of
information on events ranging from breaking news to very niche
occurrences. This has contributed even further to the growth of
Twitter's user base, and has engendered the establishment of Twitter
as a preeminent data source for users' queries -- especially about hot
topics -- on the web. In a logical extension of this phenomenon,
search engines and online retailers now consider real-time trends from
tweets in their ranking of products, dissemination of news and in
providing recommendations~\cite{abel2011analyzing,dong2010time} --
leading to large-scale pecuniary implications. However, these monetary
implications lead to increased incentives for abusing and
circumventing the system, and this is manifested as microblog
spamming. The open nature of Twitter proves to be a double-edged sword
in such scenarios, and leaves it extremely vulnerable to the
propagation of false information from profit-seeking and malicious
users (\emph{cf.} ~\cite{nytimestwitter,spamTwitterEconomist,spamTwitter}).

\begin{figure}[t]
\centering
          \includegraphics[width=.36\textwidth,trim= 0 0 0 0,clip=true
          ]{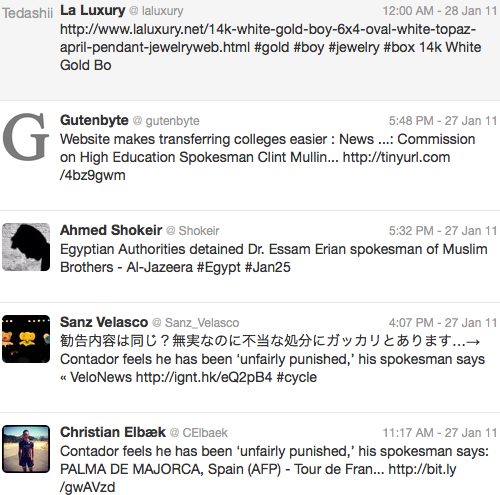}
          \vspace{-2mm}
		\caption{{\em Top-5 tweets returned by Twitter Search for the query ``White House spokesman replaced"}.}
		\label{fig:twitter-search-eg}
		          \vspace{-6mm}
\end{figure}

Unfortunately, Twitter's native search does not seem to consider the
possibility of users crafting malicious tweets, and instead only
considers the presence of query keywords in, and the temporal
proximity (recency) of, tweets~\cite{rankingTwitter}. Current Twitter search considers the recency of the tweet to be the single most important metric for judging relevance. Hence, Twitter search sorts the tweets that contain one or more query keywords by the recency of the tweet. Although, we believe recency of a tweet may be
an indicator of relevance(a tweet in the last couple of hours may be
more relevant than a tweet a week old), they may not be the sole
relevance metric for ranking. For example, for a query ``\texttt{White House spokesman replaced}" the top-$5$ tweets returned by Twitter
Search are as shown in Figure~\ref{fig:twitter-search-eg}. The tweets
are the most recent tweets at the query time and contain one or more
of the query terms. We notice that none of these five results seem to
be particularly relevant to the query.

\begin{figure}[t]
\centering
          \includegraphics[width=.4\textwidth,trim= 0 0 0 0,clip=true]{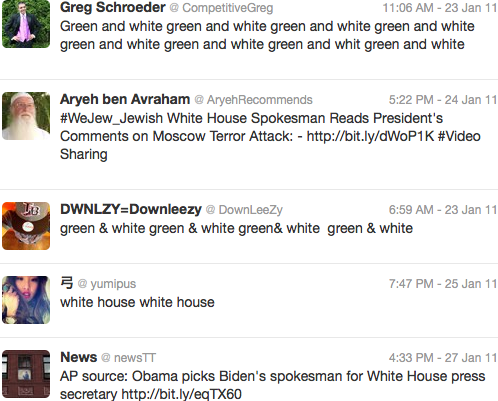}
                    \vspace{-2mm}
		\caption{{\em Top-5 tweets ranked by TF-IDF Similarity for the query ``White House spokesman replaced"}}
		\label{fig:tfidf-search-eg}
		\vspace{-5mm}
\end{figure}

Straightforward improvements such as adapting TF-IDF ranking to
Twitter unfortunately do not improve the ranking. 
Figure~\ref{fig:tfidf-search-eg} shows the results on the example
query above, but with TF-IDF ranking. In twitter, it is common to find tweets with just the query terms, with no other useful context or information. TF-IDF similarity fails to penalize these tweets. A closer inspection shows that the only relevant tweet ($5^{th}$ tweet) is from a credible news source which points to a web page that
is also trustworthy. Thus, the user/web features of a tweet may be
considered equally important as the query similarity in order to
determine the relevance of the query. We believe that ranking based on
just the user/web/tweet features results in ranking tweets that are
from trustworthy sources but may have no relation to the query.


\subsection{Our Method: RAProp}

We believe that to improve the ranking of Tweets, we must take into account the trustworthiness of tweets as well. Although Twitter supposedly considers the number of re-tweets of a tweet in its ranking, we argue that this is not sufficient--after all trustworthiness of a tweet can come not only from the trustworthiness
of the source, but also from the independent corroboration of the
content. In particular, a tweet which is independently corroborated by
many sources may well be more trustworthy than a malicious or hijacked tweet from
an otherwise trusted source. The recent multi-billion market slump sparked by hoax tweets from a hacked news paper account is an indication of impact of a false tweet
from a generally trustworthy users~\cite{APHackTweet}. Most of the current work on ranking tweets~\cite{5616236,duan2010empirical,jiang2012best}, unlike us, ignores the content of the tweet and tries to access relevance and trustworthiness from the features of the tweet and the user. These methods would consider such hoax tweets are trustworthy and relevant.

Our method, \emph{RAPRop} combines two orthogonal measures\begin{wrapfigure}{r}
  {0.20\textwidth}
  \includegraphics[width=.20\textwidth,trim= 0 0 0 0,clip=true
  ]{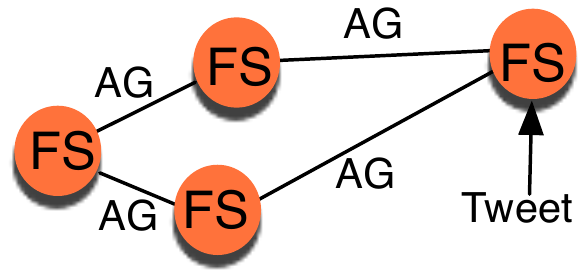} \vspace{-5mm}
		\caption{\emph{Propagation of Feature Sores(FS) over Agreement Graph (AG)}.}
		\label{fig:raprop-fig}
                \vspace{-3mm}
\end{wrapfigure}of relevance and trustworthiness of a tweet. The first, called Feature Score, measures the trustworthiness of the {\em source} of the tweet. This
is done by extracting features from a 3-layer twitter ecosystem,
consisting of users, tweets and the pages referred to in the
tweets. The second measure, called agreement analysis, estimates the
trustworthiness of the {\em content} of the tweet, by analyzing how
and whether the content is independently corroborated by other
tweets. We view the candidate result set of tweets as the vertices of
a graph, with the edges measuring the estimated agreement between each pair of tweets. The feature score is propagated over this agreement graph to compute the top-k tweets that have both trustworthy sources and independent corroboration.  
We evaluate \emph{RAPRop} on the TREC 2011 Microblog Dataset of 16~million tweets
where we compare our method against various internal baselines as well
as external baselines including Twitter Search and current best performing method in the dataset (USC/ISI). Our experiments show that \emph{RAPRop} gets a top-$30$ precision improvement of $53\%$ over current best performing method on the dataset.




In the next section, we explain how we use the user/web and tweet
features to formulate with a Feature Score for each tweet. We explain
in Section~\ref{sec:agreement} how we measure the popularity of a
topic using pairwise Agreement. In Section~\ref{sec:ranking}, we
explain how we rank our tweets which uses the Feature Score and
agreement graph generated via the methods in the preceding section. We
then discuss alternative approaches to ranking and baselines
considered in Section~\ref{sec:other-methods}. Section~\ref{sec:evaluation} presents
our evaluation. We conclude with an overview of related work.

\vspace{-2mm}
\section{Feature Score}
\label{sec:feature-score}

\begin{figure}[t]
\centering
 	\includegraphics[width=.4\textwidth, trim=0mm 0mm 0mm 0mm,clip=true]{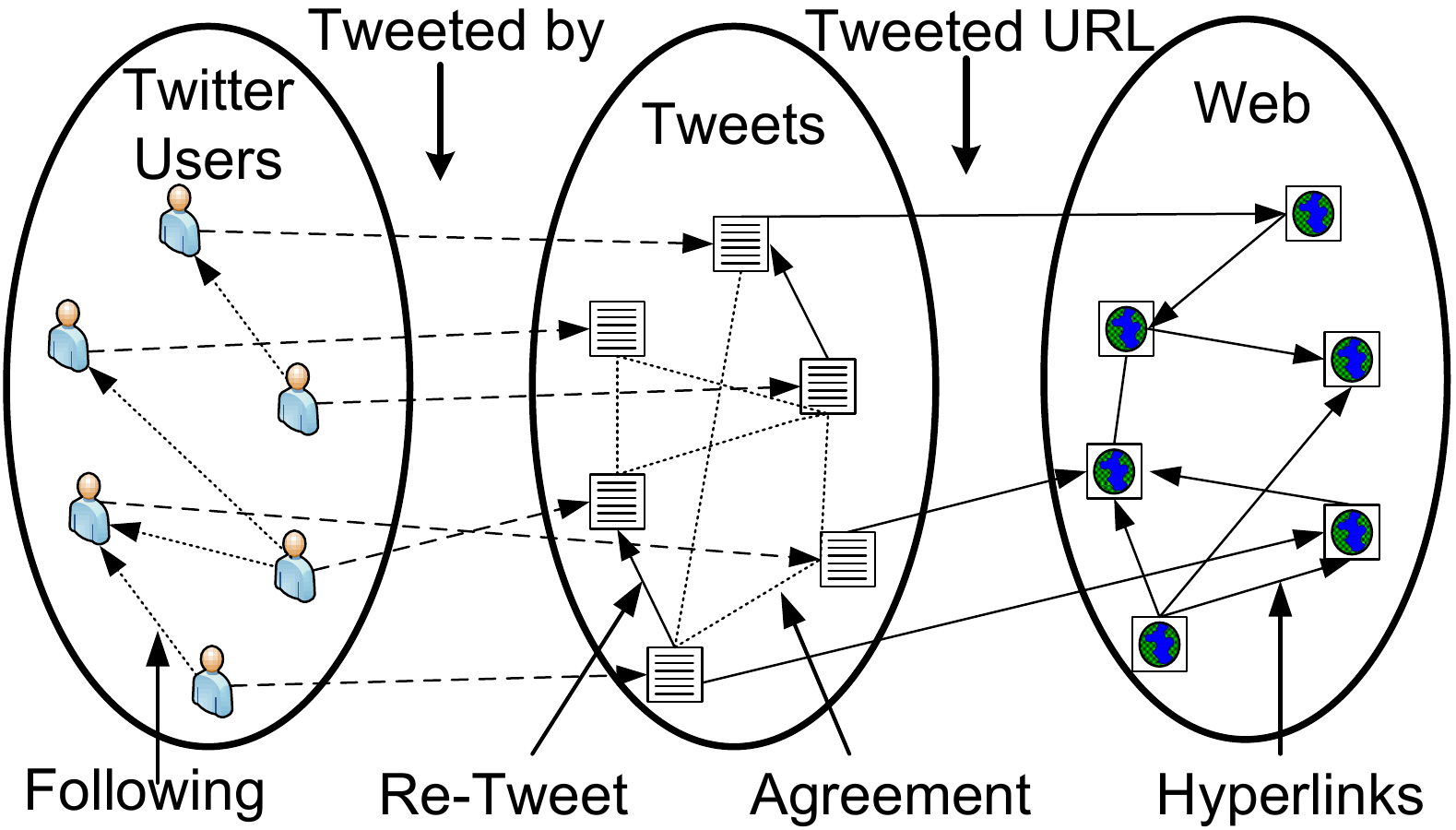}
 	          \vspace{-2mm}
 	\caption{{\em Three layer ecosystem of Twitter space composed of user layer, tweets layer and the web layer}}
 	\label{fig:tweetModel}
 	          \vspace{-5mm}
\end{figure}

In order to compute the trustworthiness of a source of a tweet, we model the entire tweet ecosystem as a three layer graph as shown in
Figure~\ref{fig:tweetModel}. Each layer in this model corresponds to
one of the characteristics of a tweet mentioned above -- the content,
the user, and the links that are part of that tweet. The user layer
consists of the set $U$ of all users $u$ such that a tweet
$t_u$ by the user $u$ is returned as part of the candidate result set $R$ for the query. Since the user base of twitter is growing exponentially, we believe that our user trustworthiness algorithm needs a high predictability of the trustworthiness of unseen users profiles. Hence, instead of computing user trustworthiness score from the follower-followee graph~\cite{cha2010measuring,yamaguchi2010turank}, we compute the trustworthiness of a user from the user profile information. The user features that we use are: \emph{follower count, friends count, whether that user (profile) is verified, the time since the profile was created, and the total number of statuses (tweets) posted by that user}. Another advantage of computing trustworthiness of a user from the user profile features is that we would be able to adjust our trustworthiness score of a user in accordance with any changes that happen in the profile(e.g.. sudden increase in the number of followers) more quickly.

The tweet layer consists of the content of the tweets in $R$; i.e.,
the tweets themselves. We select some features of a tweet that were
found to do well in determining the trustworthiness of that
tweet~\cite{infocredibility}. The features we pick include: \emph{whether the
tweet is a re-tweet; the number of hash-tags; the length of the tweet;
whether tweet mentions a user; the number of favorites received; the
number of re-tweets received; and whether the tweet contains a question
mark, exclamation mark, smile or frown}. We believe that these features are a good indicator of the trustworthiness and relevance of content of the tweet. For example the presence of a smiley or a question mark in the tweet is a good indicator the tweet is not an authoritative account on that query topic. Hence the user may not be interested in such a tweet for that query and there by making it an indicator of relevance as well. To these features, we add a feature of our own: TF-IDF similarity which is weighed by proximity of the query keywords in the tweet. Although we recognize that TF-IDF similarity may not be the sole indicator of tweet relevance to the query, we believe that a tweet that contain most of the query term may be more relevant to the query than a tweet that contain only one of the query term. Hence, these features may be considered as an indicator of the tweet's relevance to the query. Proximity of the query keywords in the tweet is a very important feature when judging the relevance. This is because we cannot rely on the mere existence of the query keywords; most tweets returned by the Twitter search interface already contain all the keywords in the query. We try to account for this in our TF-IDF similarity score by exponentially decaying the TF-IDF similarity based on the proximity of the query terms in the
tweet. Thus the similarity of a tweet $r$ to the query $Q$ is defined
as:

\vspace{-3mm}
\begin{align*}
S = \mathrm{T(t_i,Q)} \times e^{ \frac{- w \times d}{l}}
\end{align*}
\vspace{-4mm}

where $T(t_i,Q)$ is the TF-IDF similarity of the tweet $t_i$ to the query,$Q$, $w = 0.2$ is a constant(empirically decided) that decides the weight for proximity
score, $l$ is number of terms in the query and $d$ is the sum of distances between each term in the query to its nearest neighbor.

The link layer consists of the links that are used in tweets. A number of tweets link to external websites, and it would be remiss to throw that information away when considering the trustworthiness of tweets. The web has an existing, easily
query-able repository that scores web pages based on some notion of
trust and influence -- PageRank. For each tweet that contains a web
link, we instantiate a node that represents that link in the web layer
of the graph. There are links from that tweet to the node in the web
layer, as well as intra-layer links among the nodes in the web layer
based on link relationships on the open web.

The proposed ranking is performed in the tweets layer, but all three
layers are used to compute what we call the \emph{Feature Score}. The features from the user and the web page are linked to the tweets by the ``Tweeted by" relation and ``Tweeted URL" relation.


\subsection{Computing Feature Score}
\label{subsec:featurescorecompute}

Feature Score of a tweet is a measure of trust and popularity of a tweet. The popularity of a tweet may be measured by the favorites and re-tweets that tweet received, and the popularity of the user who tweeted it. The trust of a tweet comes from the user trustworthiness and trustworthiness of the web page cited in the tweet. We use the user, web page and the tweet meta information as features to compute the Feature Score.

To learn the Feature Score from features, we use a Random Forest based learning~\cite{breiman2001random} to rank method. Random Forest is an ensemble learning based classifier that creates multiple decision forests on training time using the bagging approach. We train the Random Forest with the User, Tweet and Web features described previously. We used the gold standard relevance values (described in Section~\ref{subsec:dataset-experiments}) for training and testing our model.  $5\%$ of the gold standard dataset was randomly pick
for training the model, and another $5\%$ to test the trained model
(the remaining data is reserved for the experiments). Since we did not
want to penalize tweets that do not contain a URL, or user information
that we were not able to crawl, we impute the missing feature values with population average. We normalize the Feature Score to lie between 0 and 1. Using the features chosen by this method, we get a score --- the Feature Score --- for each tweet.

\begin{figure}[t]
 	\centering
 	\includegraphics[width=.36\textwidth, trim=0 0 0 0,
  clip=true]{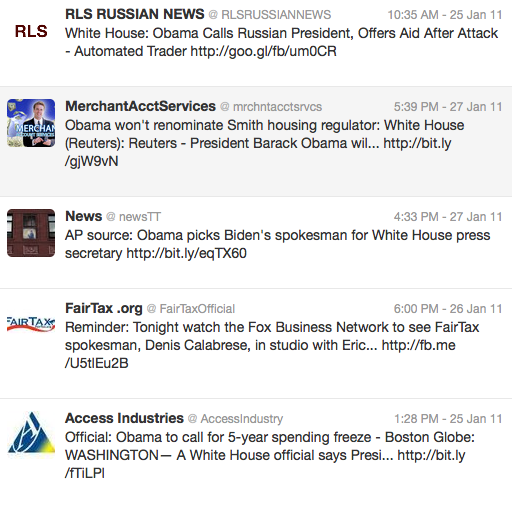}
  \vspace{-5mm}
 	\caption{{\em Top-5 tweets ranked by Feature Score(FS) for the query ``White House spokesman replaced"}}
 	\label{fig:fs-search-eg}
 	\vspace{-5mm}
\end{figure}

Since, Feature Score has been trained on features which include the trustworthiness of the user and web page and the relevance of the tweet to the query ranking considering it may be considered as a method to rank tweets considering relevance and trust. Hence, we look back again at our example query, ``\texttt{White House spokesman replaced}" results ranked using just Feature Score in Figure~\ref{fig:fs-search-eg}. We notice that the top-$5$ for the query seem to be from more reputed users and they also contain most of the query terms in them making the TF-IDF similarity to be high as well. But we also notice that in the top-$5$ results, only one tweet still seems to be relevant to the query and rest of the tweets are about other topics that just contain part of the query terms. Among the multiple topics that may exist in the candidate set of tweets, $R_Q$ returned for a query, $Q$ the user may be interested in the most popular topic. The tweets from these topics may be considered as relevant to the query than other tweets from less popular topics. On considering the trust aspect of the ranking, ranking based on just the Feature Score may lead to hoax news from trustworthy accounts may be ranked high the results due to the user popularity and trustworthiness~\cite{APHackTweet}. Hence we hypothesis that its better to rely on larger pool of independent, reasonably trustworthy users rather than relying on a single user who is highly trustworthy.

In the next section we look into how to find the tweets that is tweeted by a large pool of independent trustworthy users and there by their semantic content being popular as well as trustworthy.


\section{Agreement}
\label{sec:agreement}
Feature Score may be considered to be more of a measure of trustworthiness of the user/web page and popularity of the tweet rather than the trustworthiness of the content of the tweet. We hypothesize that a tweet on a popular topic may be relevant and trustworthy. As the popularity of a tweet is measured by the number of re-tweets it gets, the popularity of the tweet's content may be measured by the number of independent trustworthy users who endorse that content. Although the re-tweet relations among Twitter messages can be seen as an endorsement, they fall far short both because of their sparsity and that they do not capture the topic popularity rather just the tweet popularity. In this section, we develop a complementary endorsement structure among tweets by interpreting mutual agreement between two tweets as an implicit endorsement.


\subsection{Agreement as a Metric for \\ Popularity \& Trust}

Given the scale of Twitter, it is quite normal for a set of tweets returned for a query to contain tweets about multiple topics. The user is likely to be interested in only a few topics of these. Due to the temporal nature of Twitter~\cite{teevan2011twittersearch}, we hypothesize that the most popular topic is more likely to be about the breaking news that the user is interested in. Hence, the tweet from the popular topic is likely to be relevant to the user. We use the pair-wise agreement as votes in order to measure the topic popularity. Using agreement as a metric to measure popularity of a topic may be seen as a logical extension of using re-tweets to measure the popularity of a tweet. This kind of high degree of similarity can be computed from the pair-wise agreement of the content of two tweets, and this gives us a good way to measure the popularity of a tweet in terms of the number of other tweets that seem to be close to it.


Using agreement to measure the trustworthiness has been found to perform well~\cite{sourcerank} in the deep web. If two independent users agree on the
same fact -- that is, they tweet the same thing -- it is likely
that those tweets are trustworthy. As the number of users who tweet
semantically similar tweets increases, so does the belief in the idea
that those tweets are all trustworthy.



\subsection{Agreement Computation}
\label{subsec:agreementComputation}

Computing the pair-wise semantic agreement (as outlined above) between tweets at query-time, while still satisfying timing and efficiency concerns, is a
challenging task. Due to this, only computationally simple methods may
be realistically used. TF-IDF similarity has been found to perform well when measuring semantic similarity for named entity matching\cite{cohen2003comparison} and for computing semantic similarity between web database entities~\cite{sourcerank}. 
In the web scenarios, the IDF makes sure that more common words such as verbs are weighted lower than nouns which are less frequent. But due to the sparsity of verbs and other stop words in tweets, we noticed that IDF for some verbs tents to be much higher than the nouns and adverbs. Hence, we weight the TF-IDF similarity for each part of speech differently the intent is to weigh the tags that are important for agreement higher than other tags which does not highly correlate to agreement. We use a Twitter POS tagger~\cite{gimpel2011part} to identify the parts of speech of each tweet. The agreement of a pair of tweet $T_1$,$T_2$ is defined as:

\vspace{-4mm}
\begin{align*}
AG(T_1,T_2) = \sum\limits_{t\in(T_1 \cap T_2)} TF(t_1) \times TF(t_2) \times {IDF(t)}^2 \times P(t)
\end{align*}
\vspace{-3mm}

where $P(t)$ is set by us manually such that we give higher weights to POS that determines that the tweets are about the same topic such as URL($8.0$), Hashtags($6.0$), Proper noun($4.0$), Common noun/Adjective/Adverb($3.0$) and lesser weights to other POS that are less indicative of the agreement between the tweets such as Numerical($2.0$), Pronoun / Verb($1.0$), Interjection / Preposition($.5$), Existential($.2$).

%
%

We compute TF-IDF similarity on the stop word removed and stemmed
candidate set, $R_Q$.  However, due to the way Twitter's native search (and
hence our method, which tries to improve it) is set up, every single
result $r \in R_Q$ contains one or more of the query terms in $Q$. Thus the actual
content that is used for the agreement computation -- and thus ranking
-- is actually the {\em residual content} of a tweet. The residual
content is that part of a tweet $r \in R$ which does not contain the
query $Q$; that is, $r \setminus Q$. This ensures that the IDF value
of the query term as well as other common words that are not stop
words is negligible in the similarity computation, and guarantees that
the agreement computation is not affected by this. Instead of normalizing the TF-IDF similarity by the normalization factor, we divide the TF-IDF similarity only by the highest TF value. Normalization was a necessity on web where web pages have no length limit and normalization helps the web search engines penalize documents with large number of terms along with the query terms and give higher score to documents that have only fewer terms. But in the case of twitter, the document size is bound ($140$ characters). Hence we do not penalize for using the entire $140$ characters as they might bring in more content relevant to the query. We penalize tweets that repeat the terms multiple times as existence of the same term that they agree up on multiple times does not increase the agreement value. 

Agreement computation using POS weighted TF-IDF similarity may have a \emph{False Positive} if the pair of tweets is syntactically similar where as they are semantically distinct. An example of this may be the pair of tweets \emph{``BBC News: Indonesia cuts the internet"} and \emph{``BBC News cuts internet staff"} for a query \emph{BBC News staff cuts}. Since their similarity is on the query terms, the agreement score is expected to be low. This is due to the reason that IDF of the query terms are expected to be low (IDF is computed on the result set $R_Q$). There may be \emph{False Negatives} in agreement with pair of tweets that are syntactically different but semantically the same. More sophisticated approaches such as Paraphrase Detection~\cite{socher2011dynamic} or agreement computation considering synonyms from Wordnet may be considered. As these methods are much more computationally expensive than our current method we stick to POS weighted TF-IDF similarity for agreement computation. Additionally, our preliminary experiments showed that the occurrence of these false negatives are minimal.


Agreement alone may be considered to measure the trustworthiness of a document's content by measuring the number of independent users who agree with its content~\cite{sourcerank}. But we use agreement as a measure to find document(tweets) clusters that talk about the same content.  The largest cluster that is also trustworthy is likely to be the cluster that talks about the breaking news. We use the Feature Score to find the trustworthiness of the cluster and Agreement to find the size of the cluster. In the next section, we explain how we combine these orthogonal parameters by propagating the Feature Score over the agreement graph.

\section{Ranking}
\label{sec:ranking}
Our ranking of the candidate set $R_Q$ should be sensitive to: (1) relevance of a specific result $r \in R_Q$ to $Q$ by capturing the tweets about the breaking news for that query; and (2) the trust reposed in $r$. These two (at times orthogonal) metrics must be combined into a single {\em score} for each $r$, in order to make the ranking process easier. We noticed that Feature Score alone may not be the sole indicator of relevance of a tweet to the query. We believe that tweets that are part of the topic that have high content popularity (Agreement) may be more relevant to the query. But high content popularity or agreement may not be considered as the sole metric for ranking. An endorsement from a less reputed tweet (reputed user/web or popular tweet) may not be considered as equal to an endorsement from a very reputed tweet. We use the Feature Score in-order to measure the trustworthiness and popularity of a tweet. Thus, an endorsement from a tweet with higher Feature Score may be considered to be of higher value than from a lower Feature Score tweet endorsement. We compute this weighted endorsement by propagating the Feature Score over the Agreement graph to get trust-informed popularity assessment. \emph{RAProp} ranks the tweets according to these weighted endorsements. We explain the construction of the Agreement Graph, and the propagation of the Feature Score over it, in Section~\ref{subsec:agreementgraph}. The agreement graph is constructed over a set of candidate set of tweets, $R_Q$ that contain the one or more of the keywords of the query, $Q$. We explain the selection of the this candidate result set in Section~\ref{subsec:picking-result-set}.

\subsection{Agreement Graph}
\label{subsec:agreementgraph}


Computation of pairwise agreement between a pair of two tweets represents the similarity of their content to each other, not to the query Q. Tweets which have low relevance to the query term may form cliques between them and thereby gain high agreement. This problem is well known in other fields as well, for example PageRank~\cite{baeza2005pagerank} on the web.

Hence, we are not able to exploit Agreement or Feature Score by itself to compute a trustworthy and relevant Result Set. But if we base our final ranking on Feature Score, we need to provide the tweets of unpopular users but trustworthy content with a higher Feature score that they deserve. For this, we use the agreement between the tweet as a measure of deserved Feature Score of the tweet. We propagate the Feature Score to the tweets that are trustworthy but are from less reputed users. The Feature Score propagation may also be seen as a method to find which tweets out of each agreement clusters are more trustworthy. The more trustworthy cluster is either expected to contain more tweets with a higher Feature Score or larger number of nodes with a reasonably high Feature Score. In the propagation step the tweets propagate their Feature Score to not highly reputed tweets that have high agreement with the reputed tweets.

Our candidate result set $R_Q$ (for a specific query $Q$) is constructed such that all the tweets $t \in R_Q$ already bear a primary relevance to $Q$ -- tweets are chosen for inclusion in $R_Q$ as they contain one or more keywords from the query, $Q$. We propagate the Feature Score on the agreement graph that is formed by the agreement analysis detailed above. This ensures that if there is a tweet in $R_Q$ that is highly relevant to $Q$, it will not be suppressed simply because it did not have high enough Feature Score. More formally, we claim that the Feature Score of a tweet $t \in R_Q$ will be the sum of its current Feature Score and the Feature Score of all tweets that agree with $t$ weighted by the magnitude of their agreement, i.e.

\vspace{-4mm}
\begin{align*}
 S'(Q,t_i) = S(Q,t_i) + \sum_{j \in E} w_{ij} \times S(Q,t_j)\text{
} \forall \text{ }(i,j) \in E
\end{align*}
\vspace{-3mm}

where $w_j$ is the agreement between tweet $t_i$ and $t_j$ and $E$ is the edges in agreement graph.  The result set $R_Q$ is ranked by the newly computed $S'(Q,t)$. In order to perform this computation, we create a graph such that the vertices
represents the tweets and edges between the vertices represent the
agreement between them. The tweets are ranked based on the Feature Score computed after the propagation. The propagated Feature Scores may also be seen as weighted voting of other tweets that talk about the same content. The votes are weighted by their Feature Score since a vote from a highly trustworthy and popular tweet may be considered to be of higher value than a tweet from a untrustworthy tweet.

\begin{figure}[t]
 	\centering
 	\includegraphics[width=.36\textwidth, trim=0 0 0 0,
  clip=true]{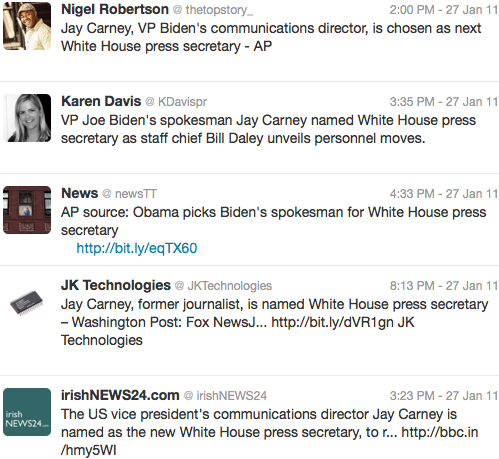}
  \vspace{-1mm}
 	\caption{{\em Top-5 tweets ranked by RAProp for the query ``White House spokesman replaced" ranked using RAProp}}
 	\label{fig:raprop-search-eg}
 	\vspace{-5mm}
\end{figure}
Our method, \emph{RAProp} ranking for the query ``\texttt{White House spokesman replaced}", achieve better results as shown in Figure~\ref{fig:raprop-search-eg}. Although the tweets in the top-$5$ results are not from very popular users and even though some of the tweets do not contain a URL, these tweets do seem to be relevant to the query as well as trustworthy in their content. The additional tweets that surfaced to the top-$5$ of the ranked results of RAProp had lesser Feature Score before propagation than the tweets in the top-$5$ of the Feature Score ranked results. The top tweets from RAProp formed a tight cluster in the agreement graph due to the fact that there were a good number of tweets that were talking about the breaking news. Although the individual tweets do not have high Feature Score, the combined Feature Score of this cluster was higher than any other topic clusters formed for this query. Thus the propagation of the Feature Score over the agreement graph makes the Feature Score for tweets in this cluster be much higher than the Feature Score of individual tweets in any other cluster. Thus the tweets that had high Feature Score before the propagation(due to the popularity of the user) but had low content agreement are pushed lower in the ranked result.

Using Feature Score weighted agreement helps us counter spam cluster voting. A tweet may be considered as malicious either due to its content being malicious in nature or because it points to a web page that is malicious in nature. When the spam tweets have tweet content as malicious, the spam tweets may have high agreement with each other as they  may contain the same content. But they would have low Feature Score as they are unlikely to have a popular user/web page. Hence the propagation of low Feature Score still keeps the propagated Feature Score to be lower than other tweet clusters that have higher Feature Score. This helps us lower the ranking of malicious tweets that form a spam cluster. When a spam tweet has a malicious link but trustworthy content, it would have a low Feature Score but the tweet is expected to have agreement with trustworthy tweets. Hence the propagation step is likely to increase the Feature Score of this spam tweet. But since we sum the Feature Score of that tweet along with the propagated Feature Score, the spam tweet is unlikely to attain as much Propagated Score as other non-spam tweets in the cluster due to their initial Feature Score being high.

On the other hand, propagation helps us counter tweets that are from highly trustworthy users (and hence high Feature Score) that may be untrustworthy~\cite{APHackTweet}. As these tweets are unlikely to have high agreement with tweets from other independent users, the propagation of the Feature Score is unlikely to increase the Feature Score of this untrustworthy tweet. Where as the tweets that may be lesser Feature Score before propagation may gain higher propagated Feature Score due to their higher content agreement. This would push these tweets higher in the ranking than the untrustworthy tweet. We evaluate the performance of our method, {\em RAProp}, in our experiments in Section~\ref{sec:evaluation} and shows that it performs better than other baselines considered.

\subsection{Picking the Result Set R}
\label{subsec:picking-result-set}
For each query of our experiments, $Q^\prime$ we collect the top-$K$ results returned by Twitter. These results become our initial candidate result set, $R^\prime$. The initial candidate result set, $R^\prime$ is then filtered to remove any re-tweets or replies. We remove the re-tweets and replies from our results set as our gold standard (TREC 2011 Microblog~\cite{TRECTwitter}) considers these tweets as irrelevant to the query. As our method does not differentiate re-tweets and replies we remove these tweets as a prepossessing step.    

We add more terms to the query,$Q^\prime$ to get the expanded query,$Q$. We select the expansion terms from the initial data set,$R^\prime$. We pick the top-$5$ nouns that have the highest TF-IDF score. In order to constrain the expansion only to nouns, we run a twitter NLP parser~\cite{gimpel2011part} to Part of speech tag the tweets. The TFs of each noun is then multiplied with its IDF value to compute the TF-IDF score. The top-$5$ terms according to the TF-IDF score is added to the query. The top-$N$ tweets returned by Twitter for the expanded query becomes the  result set,$R$.

We believe that all words in the query term are not equally important. For example, stop words or verbs are much less important than the presence of a noun in the tweet. As mentioned in Section~\ref{subsec:agreementComputation}, IDF in twitter may not be able to prioritize the presence of nouns over the presence of a stop word. Hence, we compute the TF-IDF similarity of result set,$R$ by weighting the nouns higher (an order of $10$) than other word similarity. This is especially important in the case of Twitter as it contains spam tweets that use just stop words. These tweets try to match the stop words in the query in order to be part the results. We also remove tweets that contain less than $4$ terms in them as these tweets mostly only contain the query terms and no other information.


Twitter matches query terms in URL as well while returning results. Thus, we add the URL as chunks split by special characters as part of the tweet in order for agreement to account for keywords present in the URL alone.  The tweets are stripped of punctuation, determiners, coordinating conjunctions so that agreement is only over the important terms.
\section{Other Design Choices}
\label{sec:other-methods}
In the previous sections, we focused on a specific approach, {\em RAProp}--that involves computing Feature Scores using the features from the 3-layer Twitter ecosystem, and propagating the Feature Scores over the implicit inter-tweet endorsement structure in terms of  the agreement graph.  In the following, we
describe some of the more compelling variations and discuss their
relative trade-offs with respect to {\em RAProp}. We evaluate the empirical evaluation of these design choices in Section~\ref{sec:evaluation}.

\smallskip
\noindent
\textbf{Ranking Just by Feature Score (FS):} Ranking tweets based on
only features has been attempted before~\cite{duan2010empirical,jiang2012best}. we compare the performance this kind of method -- Feature Score ({\em FS}) -- in our evaluations. Such methods make the assumption that all {\em reputed}
tweets that are pertinent to the query are relevant as well. This is
not always true - the Feature Score may not capture the true
relevance of a tweet to the query. For example, for the query ``apple
jobs'', the top results as ranked by Feature Score may be about the Apple
founder, Steve Jobs. However, the query may concern a recent jobs
report that mentions Apple Computer Inc. In such cases, our approach,
which uses the Agreement Graph created using the content of the tweets, is able to capture the popularity of the topic and therefore rank tweets pertaining to the more popular topic higher than a less relevant tweet with a higher Feature Score. We shall demonstrate that our method indeed does perform better than using just the feature
scores.

\smallskip
\noindent
\textbf{Ranking Just by Agreement (AG):} Another approach to
ranking is by ranking tweets by considering only the agreement --
using a {\em voting} methodology -- where each tweet contributes to
the other tweets' trust and hence ranking. This is used in the context
of web sources by Balakrishnan et al.~\cite{sourcerank}. However, the pairwise agreement between tweets represents the similarity of their content to {\em each other}, and says nothing about the relevance of the tweets to the query
$Q$. This may lead to the formation of cliques of high agreement but
low relevance within the result set, a problem that besets other
voting methods. Agreement-based ranking is thus highly susceptible to
irrelevant or untrustworthy tweet clusters occupying the top slots in
the ranking. Our experiments confirm this, as the agreement ({\em AG})
ranking, when used alone, has lower precision compared to our
method. 

\begin{figure*}[t]
        \centering
\begin{subfigure}[b]{.32\textwidth}
    \includegraphics[width=\textwidth,trim= 0 0 0 0,clip=true]{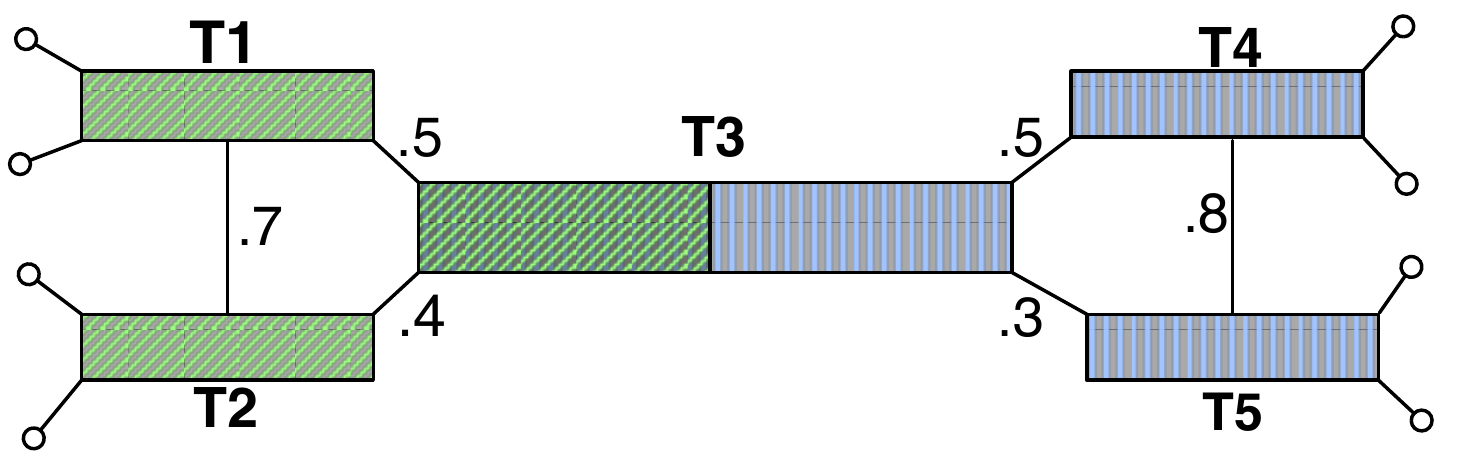}
    \caption{Illustration of propagation from trustworthy tweets
      to untrustworthy tweets}
    \label{fig:propagategraph}
\end{subfigure}
~
\begin{subfigure}[b]{.32\textwidth}
           \includegraphics[width=\textwidth]{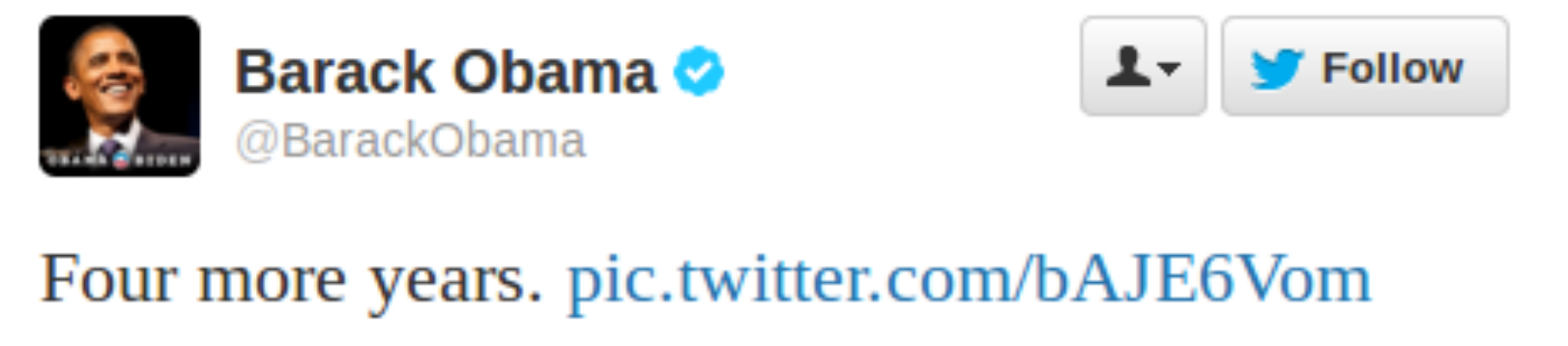}
           \caption{A trustworthy tweet with trustworthy URL}
           \label{fig:obamatweet}
\end{subfigure}
~
\begin{subfigure}[b]{.32\textwidth}
           \includegraphics[width=\textwidth]{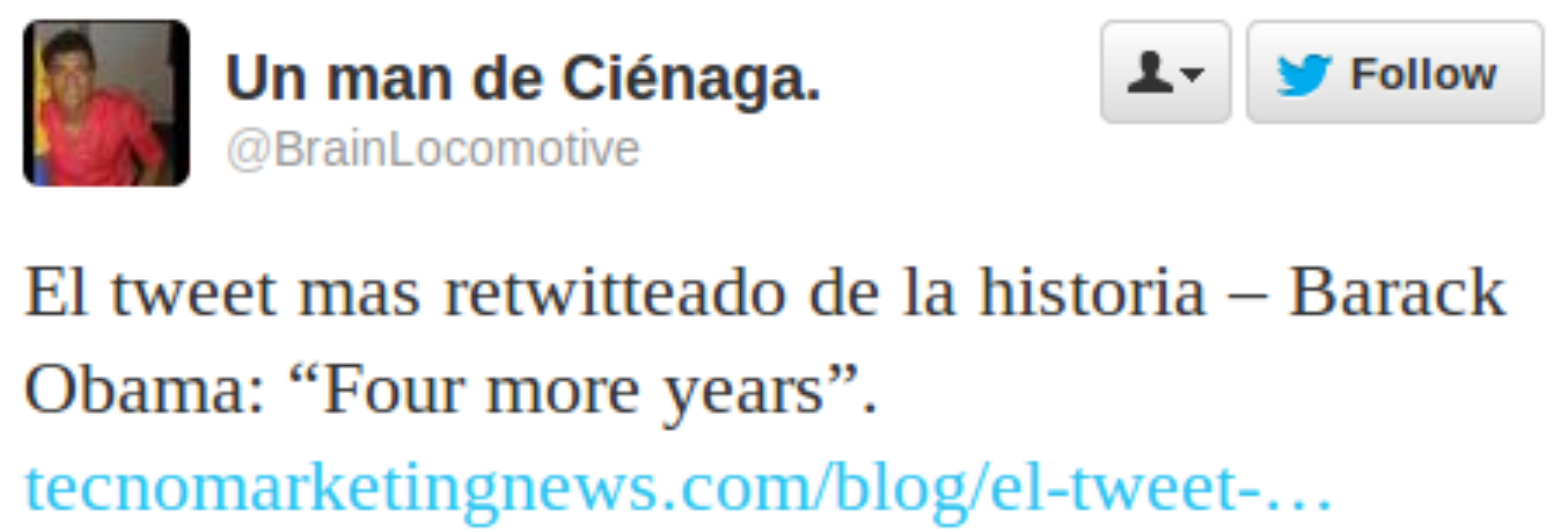}
           \caption{Spam tweet with trustworthy tweet text and a spam URL}
           \label{fig:spamtweet}
\end{subfigure}
\vspace{-2mm}
\caption{Propagation of feature score from trustworthy tweets from untrustworthy Tweets on multi-ply propagation}
\vspace{-5mm}
\end{figure*}

\smallskip
\noindent
\textbf{Ranking using Feature Score Propagation to Fix-point:} The
number of propagations of Feature Score over the agreement graph may also
be varied. Unlike other approaches of propagation of scores over graph \cite{gyongyi2004combating,brin1998anatomy}, we do not propagate our feature scores over the agreement graph until reaching a steady state. We propagate the Feature Score over the agreement graph just one-ply. 


Unlike the web scenario, the links between tweets in our case are implicit links based on agreement. Thus, for a spam tweet to get agreement with a trustworthy tweet, all it needs to do is to agree with the content of the trustworthy tweet. This is not the case in web scenario where the trustworthy user is the one who controls the explicit links in that page. 

Consider the scenario where a query on twitter, gives us the results such that there are two sets of tweets, one set contains all the tweets that contain the content which is trustworthy and the other set contains all the tweets that are spam in nature. The agreement graph we construct would have two closed connected graph that are minimally inter-connected. Let us assume that two graphs are connected by a spam tweet that tries to be part of the top results by quoting a popular tweet of the trustworthy tweet and using rest of the tweet to input untrustworthy content as Figure~\ref{fig:propagategraph}. If we do multiple propagations over the agreement graph, the Feature Score from the trustworthy tweets (T1,T2) is propagated to the untrustworthy tweets (T4,T5) through the spam tweet, T3. Thus, multiple iterations of the propagation would lead to untrustworthy tweets to be considered as trustworthy and be ranked higher in the results than they should be. Let us illustrate the above scenario with a real example from twitter. Figure~\ref{fig:obamatweet} shows the tweet by Barack Obama (which may be T1 or T2). A spammer on seeing the popularity of the tweet and the content in the tweet, tried to capitalize on the same by trying to use the same content of the popular tweet and adding malicious content along with the same(T3) as in Figure~\ref{fig:spamtweet}. This malicious tweet may be considered as the tweet that could propagate the trustworthiness from the trustworthy tweets to the untrustworthy tweets.


%


As the Feature Score for each tweet is a measure of the trustworthiness and popularity of the tweet and the user who tweeted it, we expect T1,T2 to have higher Feature Score than T3,T4,T5. Hence we need to ensure that during the propagation of the feature scores we do not propagate the feature scores from trustworthy tweets to untrustworthy tweets. The 1-ply propagation ensures that the untrustworthy tweets get only agreement from the other untrustworthy tweets such as T4, T5. The trustworthy tweets gets agreement from other trustworthy tweets such as T1,T2. T3 gets a propagated Feature Score less than T1 and T2 due to the low initial Feature Score. 


\vspace{-2mm}
\section{Experiments}
\label{sec:evaluation}

In this section, we present an empirical evaluation of our proposed approach {\em RAProp}. We compare {\em RAProp} against various baselines and design choices outlined in Section~\ref{sec:other-methods}. We also compare our method against Twitter's native search as well as the current best performing method on the TREC 2011 Microblog Dataset (USC/ISI~\cite{metzler2011usc}). We start by describing the dataset used for in experiments in Section~\ref{subsec:dataset-experiments}. We then discuss our experimental set-up in Section~\ref{subsec:experimentsetup}, and then present results that demonstrate the merits of our approach in Section~\ref{sec:results}.

\subsection{Experimental Setup}
\label{subsec:experimentsetup}

Using the set of returned tweets $R_Q$ that corresponds to a query $Q$, we evaluate each of the ranking methods. Since our dataset is offline (due to the use of the TREC dataset and the gold standard as described above), we have no direct way of running a Twitter search over that dataset. We thus simulate Twitter search (\emph{TS}) on our dataset by sorting a copy of $R_Q$ in reverse chronological order (i.e., latest first). We also use the methods discussed in Section~\ref{sec:other-methods}, as well as our proposed {\em RAProp} method, to rank $R_Q$. We set the bag size for our learning to rank method --- Random Forest --- as $10$ and the maximum number of leaves for each tree as $20$ to avoid over-fitting to the training data.

We run our experiments in two different models: mediator model and non-mediator model. In mediator model, we assume that we do not own the Twitter Data and we access twitter data only through the Twitter Search API call. Hence the tweets in the candidate result set, $R_Q$ is the most recent $N$ tweets that contain the one or more of the query term. In non-mediator model, we assume we store the entire twitter data in-house and there by we are not restricted by Twitter relevance metric to select our candidate result set, $R_Q$. We believe mediator model is a more realistic scenario and it was adopted by TREC Microblog Track by shifting to mediator model in their $2013$ contest. But we compare non-mediator model performance of our method as our baseline from the TREC 2011 Microblog Track and other related works have assumed a non-mediator model scenario.

\begin{figure*}[t]
        \centering
	\begin{subfigure}[b]{.45\textwidth}
		\centering
        \includegraphics[width=\textwidth,trim= 22 21 20 13,clip=true]{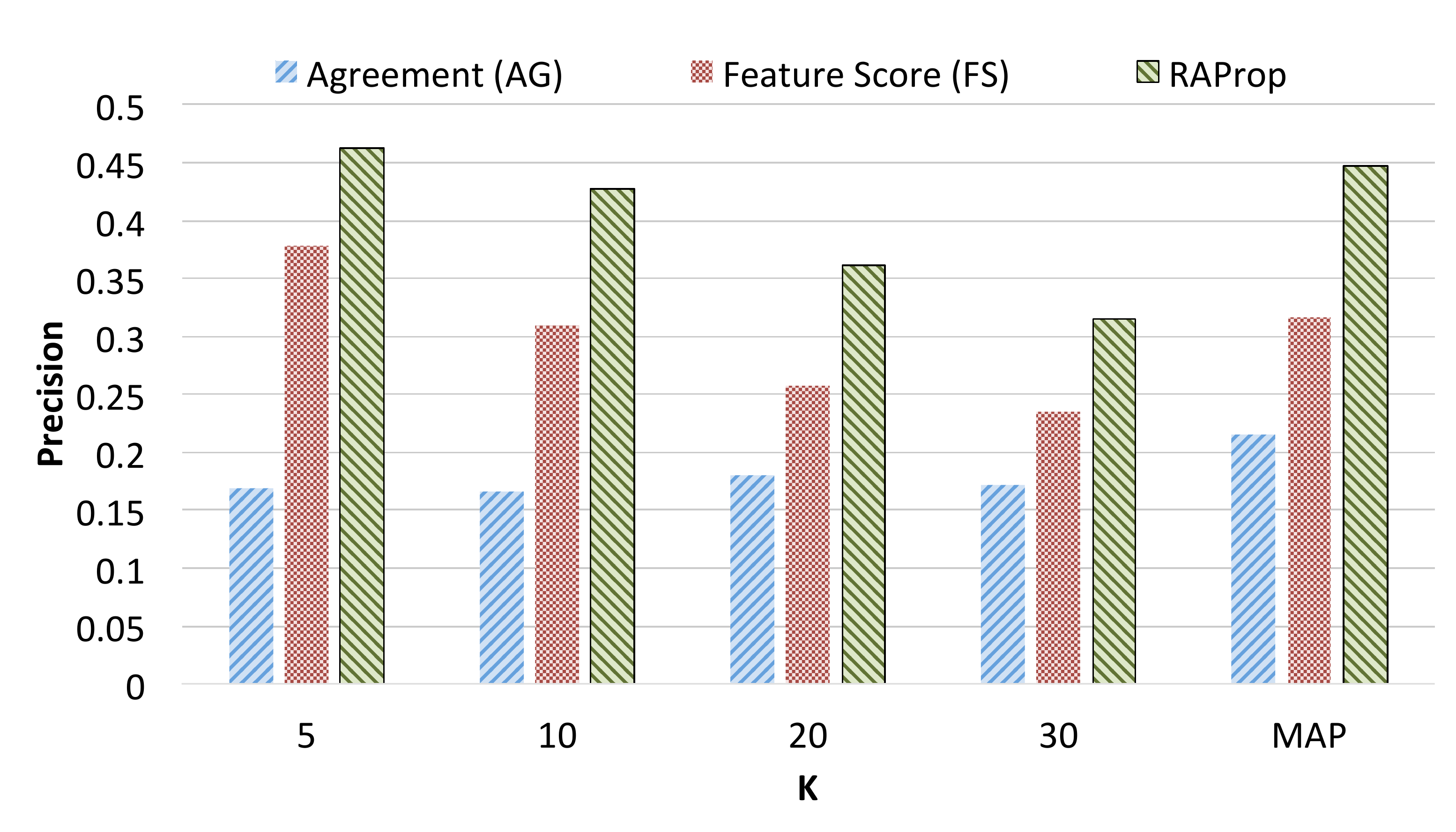}
        \vspace{-4mm}
		\caption{{\em Mediator model}}
		\label{fig:result-precision}
	\end{subfigure}
~
\vspace{-2mm}
	\begin{subfigure}[b]{.50\textwidth}
		 \centering
          \includegraphics[width=\textwidth,trim= 22 21 20 13,clip=true
          ]{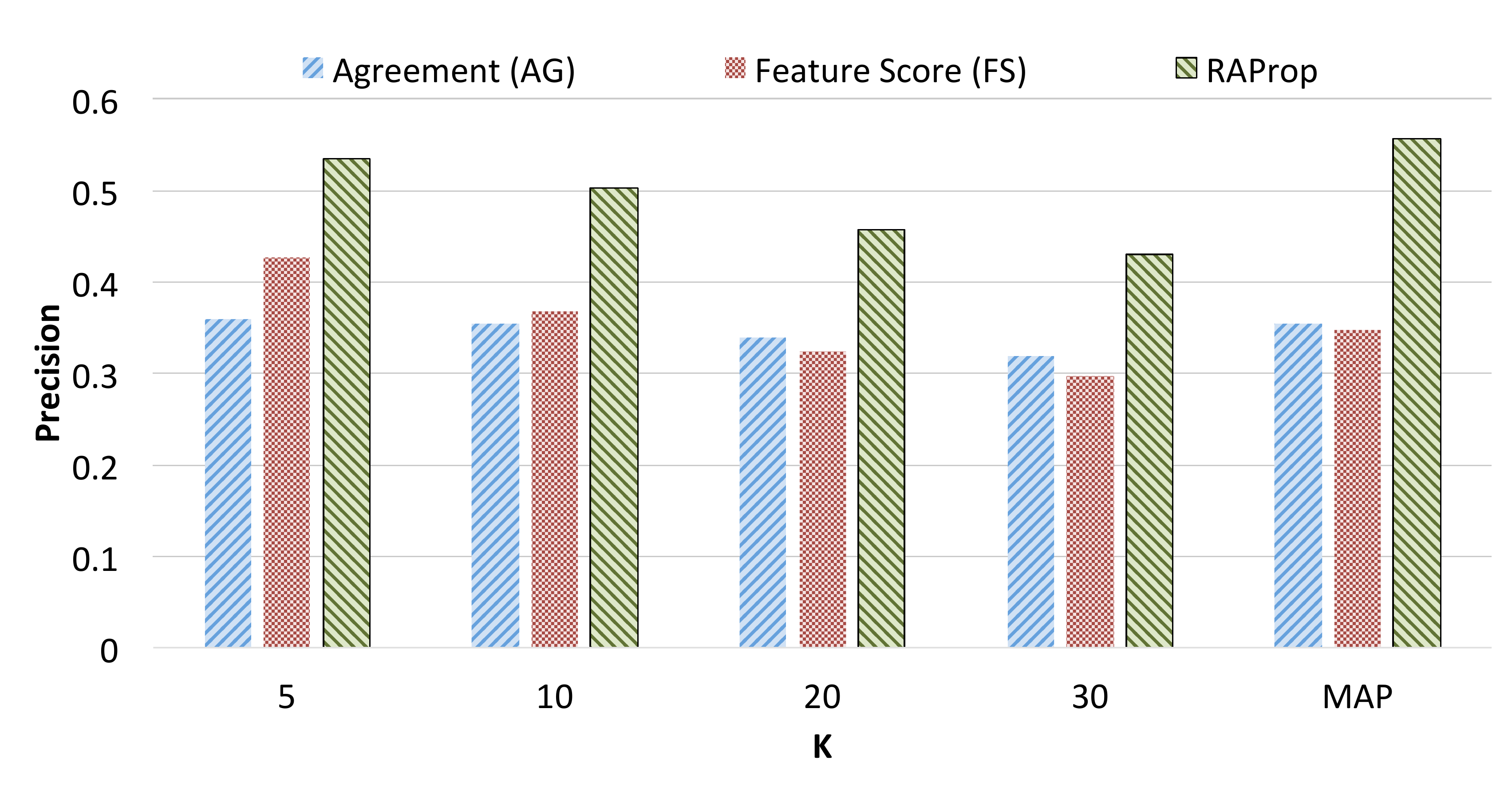}
          \vspace{-4mm}
		\caption{{\em Non-mediator model}}
		\label{fig:result-precision-nonmediator}
	\end{subfigure}
	\vspace{-1mm}
	\caption{Comparison of the proposed approach against other design choices}
	\vspace{-4mm}
\end{figure*}

\subsection{Dataset}
\label{subsec:dataset-experiments}
For our evaluation, we used the TREC 2011 Microblog Dataset~\cite{TRECTwitter}. This collection includes about 16 million tweets sampled from Twitter over a 2 week time period. It represents over 5 million micro-bloggers, at an average of 3 tweets per user. Our experiments were conducted on the 49 queries that are provided along with this dataset (and thus 49 different gold standards, one for each query, as defined previously). We used the Pagerank API in order to collect the PageRank
of all the web URLs mentioned in the tweets in this set. 

The  TREC gold standard $G_Q$ is a set of tweets annotated by TREC Microblog Track~\cite{TRECTwitter}, where the annotations are with respect to their relevance to a given query $Q$. The relevance of each tweet is denoted by 3 discrete, mutually exclusive values $\{-1,\ 0,\ 1\}$:$-1$ stands for an untrustworthy tweet, $0$ signifies irrelevance, and $1$ stands for tweets that are relevant to the query. The gold standard  gives us a way of evaluating tweets in the search results. It
is generated by humans who examine the relevance of tweets to given queries. The gold standard may be considered as a measure of trustworthiness as well, as the tweets that are marked as untrustworthy ($-1$) are considered irrelevant to the query in our evaluations.

The maximum achievable precision in this dataset for $30$ results($K = 30$) by re-ranking $R_Q$ averaged over all $49$ queries is $0.498$ while considering mediator model and $0.684$ while considering a non-mediator model. Since we are interested in the relative performance of our method against the internal and external baselines, this is not a matter of concern.

\subsection{Internal Evaluation of methods}
\label{sec:results}

We compare our method,\emph{RAProp} against the other design choices mentioned in Section~\ref{sec:other-methods}. We compare the precision of the different methods both in a mediator model as well as a non-mediator model. In the mediator model, we pick the top-$N$ tweets that our simulated twitter returns and this is the input to all the various methods. In the non-mediator model, the top-$N$ tweets is selected by the TF-IDF similarity of the tweet to the query.

\subsubsection{Internal Evaluation of methods \\ in a mediator model}

We compared the top-$K$ Precision at $5$, $10$, $20$, $30$ and MAP(Mean Average Precision), of our method, \emph{RAProp} along with the various approaches proposed in Section~\ref{subsec:experimentsetup}. Not all relevant tweets from the dataset for the query are not evaluated for its relevance to the query and may not be part of the gold standard. Since we are not sure the relevance of the tweets not part of the gold standard, we ignore those tweets that are not part of the gold standard while computing the precision value. We pick the $N$ most recent tweets that contain one or more of the query keywords. For our experiments we set $N = 2000$.


Figure~\ref{fig:result-precision}, supports our hypothesis that {\em RAProp} has better precision values than using Feature Score alone (\emph{FS}) or Agreement (\emph{AG}) alone for ranking. Since there exist less than $K$ relevant documents in the Result Set $R$, the precision values are expected to drop as the value of $k$
increases. However, {\em RAProp} maintains its dominance over the other methods and the baseline and achieves a $34\%$ improvement at Precision at $30$ results over the next highest performing method, FS. Additionally, an improved of $41\%$ of {\em RAProp} over Feature Score show that \emph{RAProp} is able to place relevant results higher when compared to the other methods. 


\subsubsection{Internal Evaluation of methods \\ in a non-mediator model}
We compared the top-$K$ Precision at $5$, $10$, $20$, $30$ and MAP, of the proposed method assuming we have the entire twitter dataset. This allows us to choose the Result Set, $R$ from the entire data set instead of top-$N$ from simulated twitter results. We choose the Result Set, $R$ by picking the top-$N$ tweets according to TF-IDF similarity of the tweet to the query, as mentioned in Section~\ref{subsec:picking-result-set}.

As we can see from Figure~\ref{fig:result-precision-nonmediator}, our method gets better precision scores than all other design choices considered and achieves a $35\%$ improvement at Precision at $30$ results and a $57\%$ improvement in MAP over the next highest performing method, AG. This proves that our method, \emph{RAProp}, is able to achieve higher precision even on a non-mediator model where the Result Set, $R$ is expected to have higher number of relevant documents.


\begin{figure*}[t]
        \centering
	\begin{subfigure}[b]{.49\textwidth}
		\centering
        \includegraphics[width=\textwidth,trim= 22 22 30 33,clip=true]{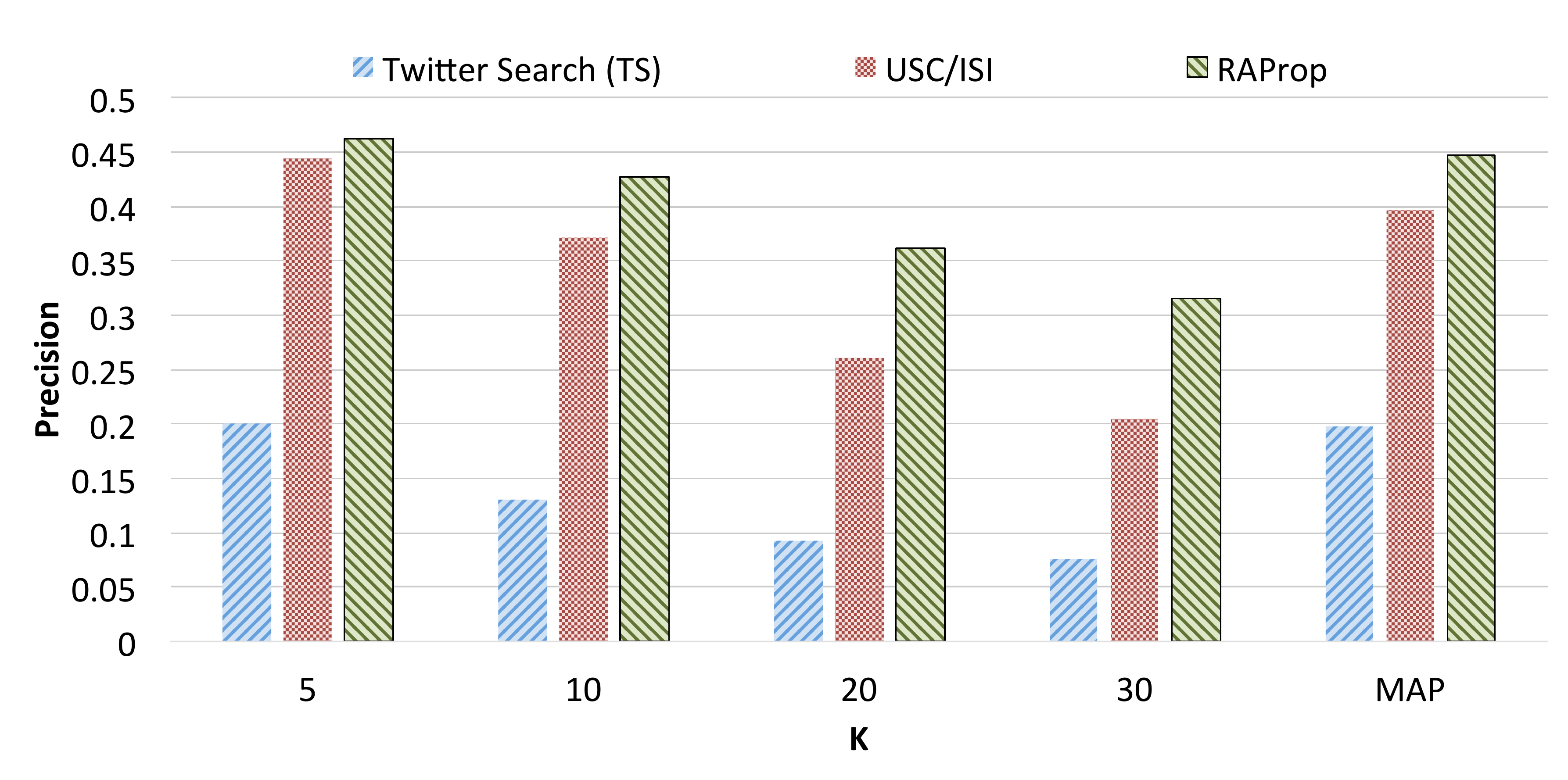}
        \vspace{-4mm}
          \caption{{\em Against Twitter and USC/ISI while assuming a mediator model}}
		\label{fig:external-mediator}
	\end{subfigure}
~
	\begin{subfigure}[b]{.49\textwidth}
		 \centering
          \includegraphics[width=\textwidth,trim= 22 20 45 33,clip=true]{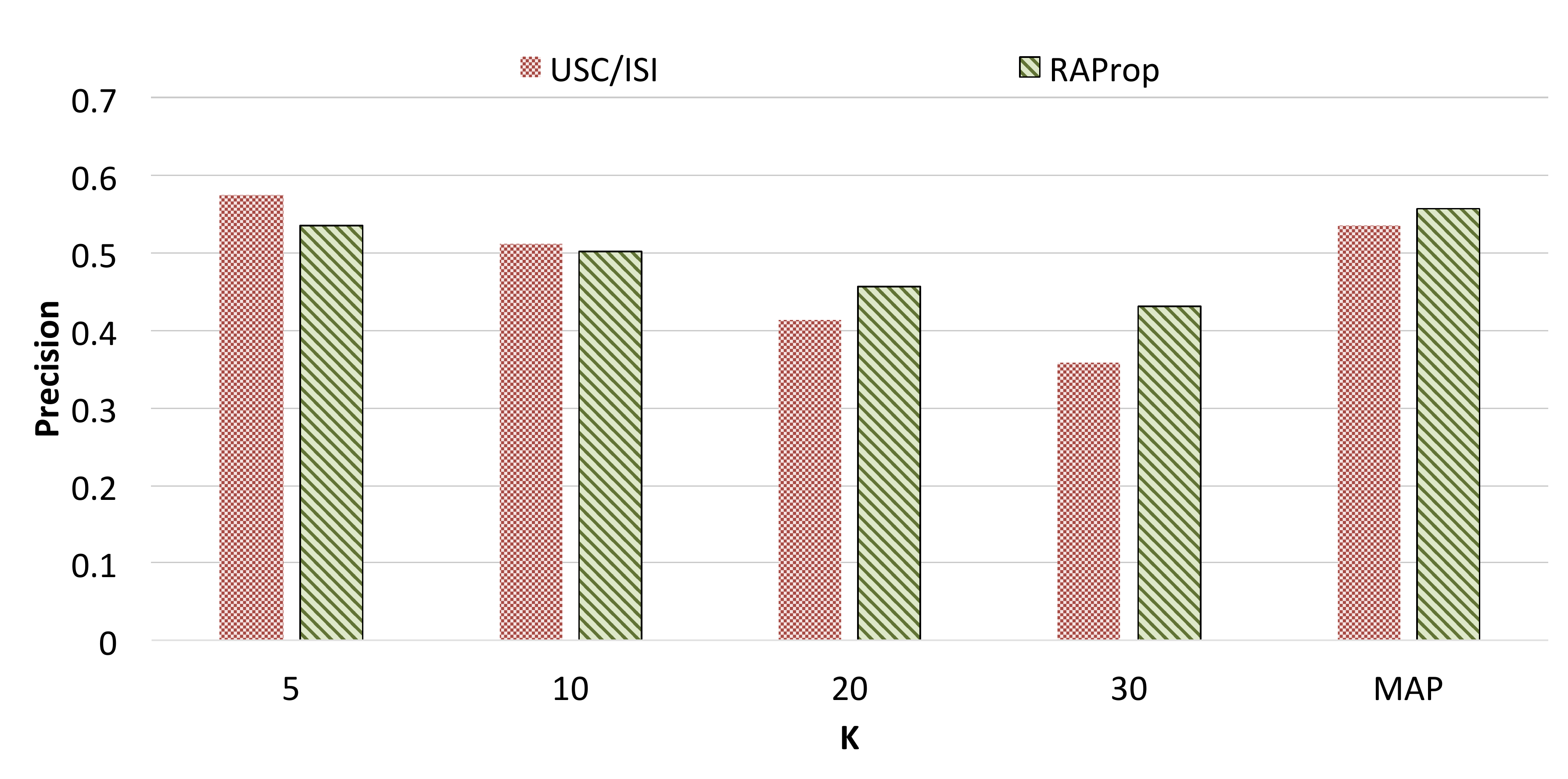}
          \vspace{-4mm}
          \caption{{\em Against USC/ISI on a non-mediator model}}
		\label{fig:external-non-mediator}
	\end{subfigure}
	\vspace{-1mm}
	\caption{External Evaluation of \emph{RAProp}}
	\vspace{-1mm}
\end{figure*}

\subsubsection{1-ply {\em vs.} Multiple Ply}
\label{subsubsec:oneply-multiply-propagate}

\begin{figure}[ht]
\centering
          \includegraphics[width=.4\textwidth,trim= 0 0 0 0,clip=true]{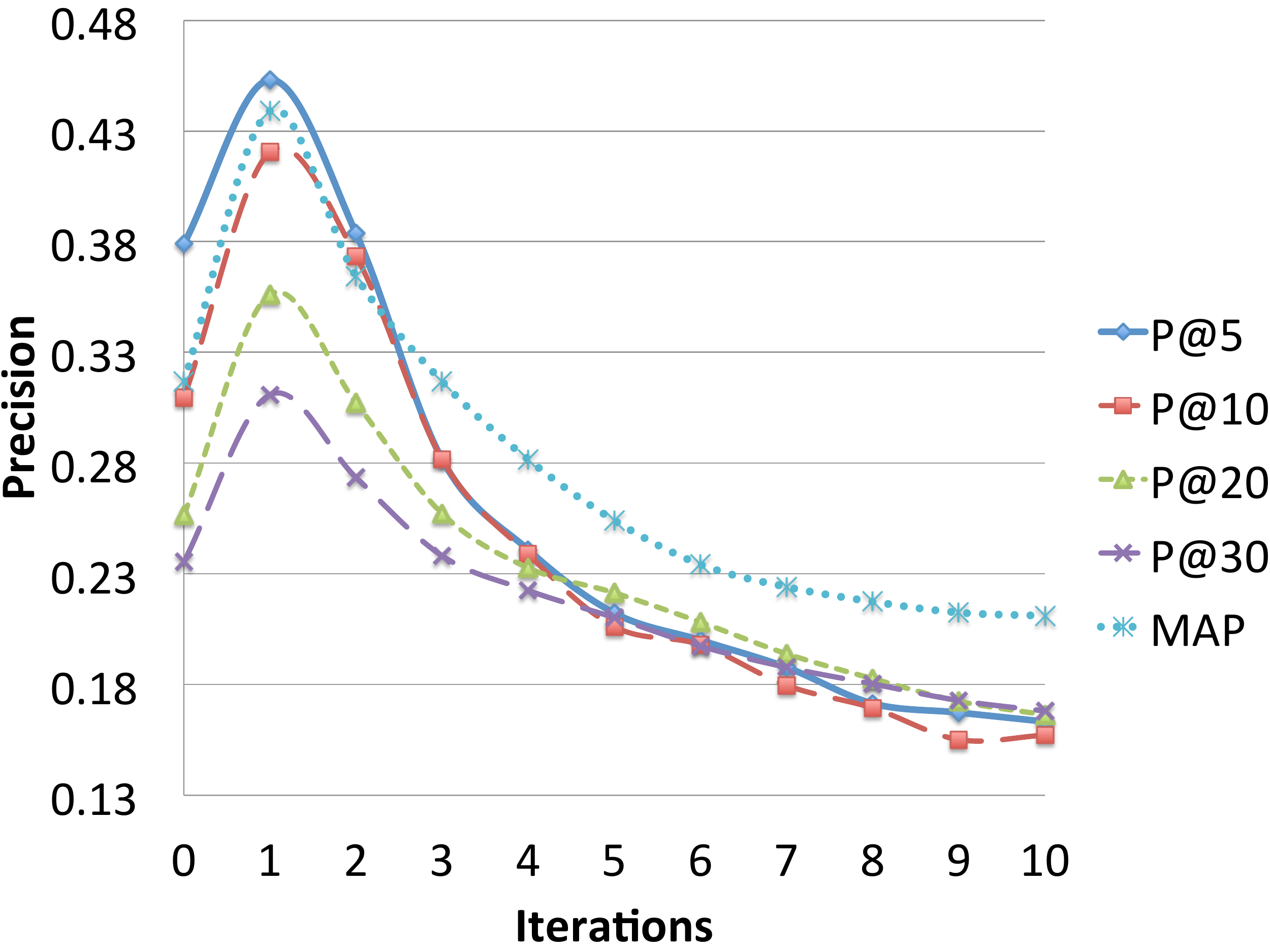}
          \vspace{-1mm}
          \caption{{\em Precision and MAP across multiple propagations of
            RAProp}}
		\label{fig:result-iterations}
		\vspace{-3mm}
\end{figure}

We compare the top-$k$ Precision at $5$, $10$, $20$, $30$ results and MAP values for various numbers of propagations over the Agreement Graph. Zero iterations can be considered as ranking based only on initial Feature Scores, which is the {\em FS} method. One iteration over the agreement graph is the {\em RAProp} method. As shown in Figure~\ref{fig:result-iterations}, propagating the Feature Score over the agreement graph certainly improves the Precision and MAP scores. However, we see that multiple iterations lead to a reduction of Precision and MAP scores. This validates our claim in Section~\ref{sec:other-methods} that multiple propagations will lead to a decrease in relevance.

\subsection{External Evaluation of methods}
\label{subsec:external-eval}


In this section, we evaluate the performance of our method \emph{RAProp} to two other external baselines, Twitter Search and USC/ISI method~\cite{metzler2011usc}. We also compare our method with the TREC Microblog 2011 best performing method by Metzler and Cai (USC/ISI)~\cite{metzler2011usc}. USC/ISI uses a full dependence Markov Random Field model, Indri, to achieve a relevance score for each tweet in the dataset. Indri creates an off-line index on the entire tweets dataset in order to provide a relevance score for each tweet in the entire tweets dataset. This score along with other tweet specific features such as tweet length, existence of a URL or a hash-tag is used by a Learning to Rank method to rank the tweets. In our experiments, we compare the performance of our method against the USC/ISI method both in a mediator and non mediator model. In the non-mediator model, we run the queries over the entire tweet dataset index. On the mediator model, since we assume we do not have access to the entire dataset, we create a per-query index on the top-$N$ tweets returned by twitter for that query.

We compare the performance of our method over these baselines while assuming a mediator model as well non-mediator model. As shown in Figure~\ref{fig:external-mediator}, when we assume a mediator model our model, \emph{RAProp} achieves higher precision for all values of K ($10$,$20$,$30$) than both current Twitter Search and USC/ISI method. When we compare the top-$30$ precision of \emph{RAProp} against USC/ISI method and Twitter Search, we achieve a $53\%$ and $300\%$ improvement respectively. \emph{RAProp} also achieves more than $125\%$ and $13\%$ higher MAP scores than Twitter search and USC/ISI method.

We also compare the precision of \emph{RAProp} against USC/ISI method in a non-mediator model. In this method, USC/ISI method is able to index the entire tweet database. The queries are run over this index and the similarity score of each tweet returned by Indri is then combined with other features to rank the tweets for that query. We then compare the top-$K$ ranked results with the results of \emph{RAProp}. As shown in Figure~\ref{fig:external-non-mediator}, we noticed that precision at K obtained by RAProp is equal to that of USC/ISI for K=$10$, and gives better results for higher values of K. \emph{RAProp} is able to achieve a $20\%$ higher top-$30$ precision than USC/ISI. Also, \emph{RAProp} achieves a $4\%$ higher MAP values than the USC/ISI ranking. This shows we are able to rank more relevant results higher in the ranking than USC/ISI ranking.


\section{Related Work}
\label{sec:related-work}
Although ranking tweets has received attention recently
(c.f. \cite{TRECTwitter,metzler2011usc}), much of it is focused
only on relevance. Most such approaches need background
information on the query term which is usually not available for
trending topics. A quality model based on the probability of
re-tweeting~\cite{qualitymodel2012trec} has been proposed which tries
to associate the words in each tweet to the re-tweeting
probability. We believe that the re-tweet probability of a tweet may not directly co-relate to the relevance of the tweet. This is because re-tweet probability of a tweet determines if the tweet is needed to be broadcast to the user's followers while relevance determines if the tweet is informative to the users these are orthogonal issues. There are also multiple approaches ~\cite{5616236,duan2010empirical,jiang2012best,
BeTaBo2012.3,cha2010measuring} that try to rank tweets based on specific features of the user who tweeted the tweet. These methods are comparable to the Feature Score ({\em FS}) method. Our approach complements these by measuring popularity of the content of the tweets by using the Feature Score as trustworthiness and popularity of the user, and can thus be seen as folding many of the features from previous work into a ranking algorithm. Ranking using the Web Page mentioned as a part of the tweet have been considered~\cite{mccreadie2012relevance}. We believe that adding web page content to the tweet dilutes the content of the tweet and hence ranking would be based solely on the content of the web page. Hence, the ranking would degrade to ranking web pages.

The user follower-followee relation graph has been used to compute the popularity and trustworthy of a user~\cite{cha2010measuring,yamaguchi2010turank,abbasi2013measuring}. These approaches have no predictability when it comes to a user who is not part of the data set on which the popularity was found. They also take much longer for a change in the relation graph to reflect in the popularity score as the algorithm needs to be run over the entire follower-followee relation graph so as to get the new popularity values. 

Credibility analysis of Twitter stories has been attempted by Castillo
et al.~\cite{infocredibility,eventcredibility}, who try to classify Twitter story
threads as credible or non-credible.  Our problem is different, since
We try to assess the credibility of individual tweets. As the feature
space is much smaller for an individual tweet -- compared to Twitter
story threads -- the problem becomes harder. 

Propagating trust over explicit links has been found to be effective in web
scenarios~\cite{brin1998anatomy,gyongyi2004combating,
artz2007survey,richardson2003trust}. We cannot apply these directly to micro-blog scenarios as there are no explicit links
between the documents.  Finding relevant and trustworthy results based
on implicit and explicit network structures has been considered
previously~\cite{gupta2011heterogeneous,sourcerank}. Real time web search considering tweet ranking has also been attempted~\cite{abel2011analyzing,dong2010time}. We consider the
inverse approach of considering the web page ``prestige'' to improve
the ranking. To the best of our knowledge, ranking of
tweets considering trust and content popularity has not been
attempted. Ranking tweets based on the propagated user authority
values have been attempted by Yang~\cite{yang2012finding}. Since the
propagation is done over the re-tweet graph, we expect tweets from
popular users to be ranked higher. In contrast, we base our ranking also on the content and relevance to the query.

\section{Conclusion}
\label{sec:conclusion}

In this paper, we proposed \emph{RAProp}, a microblog ranking
mechanism for Twitter that combines two orthogonal features of
trustworthiness--trustworthiness of source and trustworthiness of
content,  in order to filter out irrelevant results and spam. RAProp works by computing Feature Score for each tweet and propagating that over a graph that represents content-based agreement between tweets, thus leveraging the collective intelligence embedded in tweets. Our detailed experiments on a large
TREC dataset showed that RAProp improves the precision of the returned
results significantly over internal and external baselines while
considering a mediator as well as non-mediator models.

\small{
\bibliographystyle{abbrv}
\bibliography{microblog}
}
\end{document}